\documentclass[aps,prl,amsmath, amssymb, english, aps, prl, twocolumn,reprint,floatfix, superscriptaddress]{revtex4-1}

\usepackage[english]{babel}
\usepackage{graphicx}
\usepackage{verbatim}

\usepackage{epsfig}
\usepackage{epstopdf}
\usepackage{amsfonts}
\usepackage{amsthm}
\usepackage{amsmath}
\usepackage{amssymb}
\usepackage{color}
\usepackage[usenames,dvipsnames,svgnames,table]{xcolor}
\usepackage{hyperref} 
\hypersetup{backref,pdfpagemode=FullScreen,colorlinks=true,linkcolor=NavyBlue,citecolor=BrickRed}
\usepackage{makeidx}
\usepackage{pifont}

\usepackage{color}
\usepackage{grffile}


\def\expect#1{\mathinner{\langle{#1}\rangle}}

{\catcode`\|=\active
  \gdef\expect#1{\left<\mathcode`\|"8000\let|\bravert {#1}\right>}}
\def\bravert{\egroup\,\vrule\,\bgroup}

\def\beq{\begin{equation}}
\def\eeq{\end{equation}}
\def\be{\begin{equation}}
\def\ee{\end{equation}}

\def\cG0{{\cal G}_0}

\def\spinup{\uparrow}
\def\spindown{\downarrow}

\def\eps{\epsilon}

\def\g{\gamma}

\def\s{\sigma}

\def\uc2{$U_{c2}$}
\def\uc1{$U_{c1}$}

\def\bea{\begin{eqnarray}}
\def\eea{\end{eqnarray}}
\def \bal{\begin{align}}
\def \eal{\end{align}} 
\def\#{\!\!}
\def\@{\!\!\!\!}

\def\+{\dagger}

\def\up{\spinup}
\def\down{\spindown}

\begin{document}

\title{\bf Mott Quantum Critical Points at finite doping}

\author{Maria Chatzieleftheriou}
\affiliation{Laboratoire de Physique et Etude des Mat\'eriaux, UMR8213 CNRS/ESPCI/UPMC, Paris, France}

\author{Alexander Kowalski}
\affiliation{Institut f\"ur Theoretische Physik und Astrophysik and W\"urzburg-Dresden Cluster of Excellence ct.qmat, Universit\"at W\"urzburg, 97074 W\"urzburg, Germany}

\author{Maja Berovi\'{c}}
\affiliation{International School for Advanced Studies (SISSA), Via Bonomea 265, I-34136 Trieste, Italy}

\author{Adriano Amaricci}
\affiliation{CNR-IOM DEMOCRITOS, Istituto Officina dei Materiali,
Consiglio Nazionale delle Ricerche, Via Bonomea 265, I-34136 Trieste, Italy}

\author{Massimo Capone}
\affiliation{International School for Advanced Studies (SISSA), Via Bonomea 265, I-34136 Trieste, Italy}
\affiliation{CNR-IOM DEMOCRITOS, Istituto Officina dei Materiali,
Consiglio Nazionale delle Ricerche, Via Bonomea 265, I-34136 Trieste, Italy}

\author{Lorenzo De~Leo}
\affiliation{Laboratoire de Physique et Etude des Mat\'eriaux, UMR8213 CNRS/ESPCI/UPMC, Paris, France}

\author{Giorgio Sangiovanni}
\affiliation{Institut f\"ur Theoretische Physik und Astrophysik and W\"urzburg-Dresden Cluster of Excellence ct.qmat, Universit\"at W\"urzburg, 97074 W\"urzburg, Germany}

\author{Luca de'~Medici}
\affiliation{Laboratoire de Physique et Etude des Mat\'eriaux, UMR8213 CNRS/ESPCI/UPMC, Paris, France}


\maketitle


{\bf 
Strongly correlated materials often undergo a Mott metal-insulator transition, which is tipically first-order, as a function of control parameters like pressure\cite{imada_mit_review}. Upon doping, rich phase diagrams with competing instabilities are found. Yet, the conceptual link between the interaction-driven Mott transition and the finite-doping behavior\cite{Castellani-Mott_gas-liquid}\cite{Yee_Balents-PhaseSep_Mott} lacks a clear connection with the theory of critical phenomena.
In a prototypical case of a first-order Mott transition the surface associated with the equation of state for the homogeneous system is "folded" so that
in a range of parameters stable metallic and insulating phases exist and are connected by an unstable metallic branch\cite{Ono_multiorb_linearizedDMFT,Strand_fixpoint_DMFT,Tong_Hubbard_sigmoid_1storder}.
Here we show that tuning the chemical potential the zero-temperature equation of state gradually unfolds. Under general conditions, we find that the Mott transition evolves into a first-order transition between two metals, associated to a phase separation region ending in a quantum critical point (QCP) at finite doping. This scenario is here demonstrated solving a simple multi-orbital Hubbard model relevant for the Iron-based superconductors, but its origin - the splitting of the atomic ground state multiplet by a small energy scale, here Hund's coupling - is much more general. A strong analogy with cuprate superconductors is traced.}


Mott physics, charge instabilities and quantum criticality are recurrent leitmotifs in the field of strongly correlated materials.  
Their connection was explored early on theoretically for the cuprate superconductors\cite{Emery_Kivelson-PhaseSep_tJ,Grilli_RCDK-PhaseSep_pdmodel,CDG_PRL95,Imada_2D_Cuprate_QMC}. These are indeed doped Mott insulators and host both a "strange", possibly quantum critical metal, and incommensurate charge-density wave phases\cite{Arpaia_Ghiringhelli_Review}. 
Moreover charge instabilities occur in a variety of other correlated systems like e.g. titanates\cite{Zhou_Goodenough-phase-separation_Titanates} and transition-metal dichalcogenides\cite{Sipos_Forro-Dichalcogenides_phase-separation}.
However despite the great interest in this topic a clear physical picture of the conditions leading to phase separation and quantum criticality in doped Mott insulators is still missing. 

In this work we broaden the perspective and show that a phase separation zone ending in a QCP is an intrinsic features connected to the Mott transition.
We address this issue within a different framework, namely a Hund's metal which is realized in a doped multiorbital Hubbard model.
We can thus both build on the recent understanding of Hund's metals triggered by iron-based superconductors\cite{Yin_FeSC_kinetic_frustration,Werner_SpinFreezing,Ishida_Mott_d5_nFL_Fe-SC,demedici_OSM_FeSC,Georges_annrev} and attack the problem by solving a simplified model using Dynamical Mean-Field Theory (DMFT)\cite{georges_RMP_dmft} in a numerically exact way, ruling out any ambiguity connected with the numerical solution.  
More specifically we use DMFT solved by Numerical Renormalization Group (NRG) at zero temperature to study a two-orbital Hubbard model with on-site Coulomb repulsion U and Hund's exchange coupling J, which favours high-spin states on every atom\cite{Georges_annrev} (see Methods). 
\begin{figure*}[htb]
\begin{center}
\includegraphics[width=\textwidth]{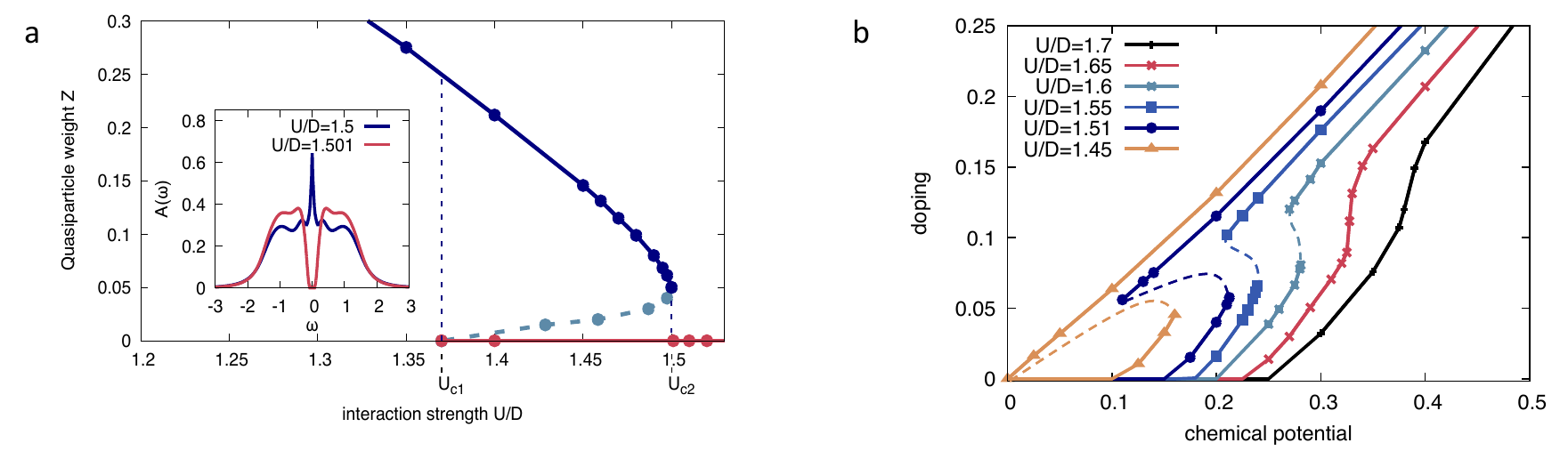}
\caption{{\bf Multiple solutions close to the Mott transition at T=0.} 
{\bf a}, Interaction-driven transition. We plot the quasiparticle weight at T=0 and half-filling in the 2-orbital Hubbard model with density-density interaction and Hund's coupling J/U=0.25, calculated within Dynamical Mean-Field Theory (DMFT) solved by Numerical Renormalization Group (NRG). The first-order character of the Mott transition is embodied by the sigmoidal shape of the curve in the range of interaction strengths U$_{c1}$$<$U$<$U$_{c2}$ where two stable solutions - one metallic at finite Z (blue) and one insulating at Z=0 (red) - coexist, and are connected by a third unstable metallic branch (light blue dashed line). Inset: change of spectral function between the metallic and insulating stable solutions when passing U$_{c2}=1.5D$ (where D is the half-bandwith of the non-interacting system). {\bf b}, Density-driven transition. We plot the density (measured as doping from half-filling) vs chemical potential curves for several values of the interaction strength U/D. The Mott insulator is incompressible and is thus indicated by the horizontal plateau at half-filling, while the doped solutions are metallic. The points are calculated within NRG-DMFT, the dashed lines sketch the unstable branch connecting the two stable ones, as deduced from calculations within ED-DMFT (see Supplementary Information). The adiabatic connection of the solutions implies the crossing of the energy between the two stable branches at some $\mu_c(U)$ in the coexistence zone, and thus a discontinuous jump from one to the other (corresponding to a Maxwell construction). The sigmoidal shape of the curves reported in both panels reflects the folding of the equation of state surface, of which a 3D visualization as a function of both U and $\mu$ is given in panels b and c of Fig.~\ref{fig:PhDiag}.}\label{fig:Z_vs_U}
\end{center}
\end{figure*}
Our study of this  simple and paradigmatic model shows that the generic first-order character of the transition for  J$\neq$0\cite{Ono_multiorb_linearizedDMFT,pruschke_Hund,Hallberg_Cornaglia-DMRG_Hund}, which implies two stable solutions, can be linked directly to finite-doping instabilities. We argue that these results are general to a wide class of models where another energy scale besides the Hubbard U is present.


We start from the half-filled system. In Fig.~\ref{fig:Z_vs_U}a we show the zero-temperature quasiparticle weight Z, which is a measure of the system's metallicity (see Methods) and whose vanishing signals the Mott transition.
As a function of the interaction strength U the metallic solution does not evolve continuously in the insulating one though, as testified by the sharp change in the spectrum: the gap opens abruptly beyond a threshold value labeled U$_{c2}$, while the central peak - of which Z is the spectral weight - disappears. Importantly, the insulating solution with Z=0 exists not only for all larger values of U but also for a range U$_{c1}$$<$U$<$U$_{c2}$ where the equation of state of the system is then multi-valued. The actual transition point U$_c$ is where the energies of the two solutions cross.

A crucial feature is that the two stable solutions are adiabatically connected\cite{Ono_multiorb_linearizedDMFT,Strand_fixpoint_DMFT,Tong_Hubbard_sigmoid_1storder} through a third, unstable metallic branch joining the stable metallic branch in U$_{c2}$ to the stable insulating branch in U$_{c1}$. This implies that, following by continuity the three solutions, the equation of state is folded into a characteristic sigmoidal shape. 

\begin{figure*}[htb]
\begin{center}
\includegraphics[width=\textwidth]{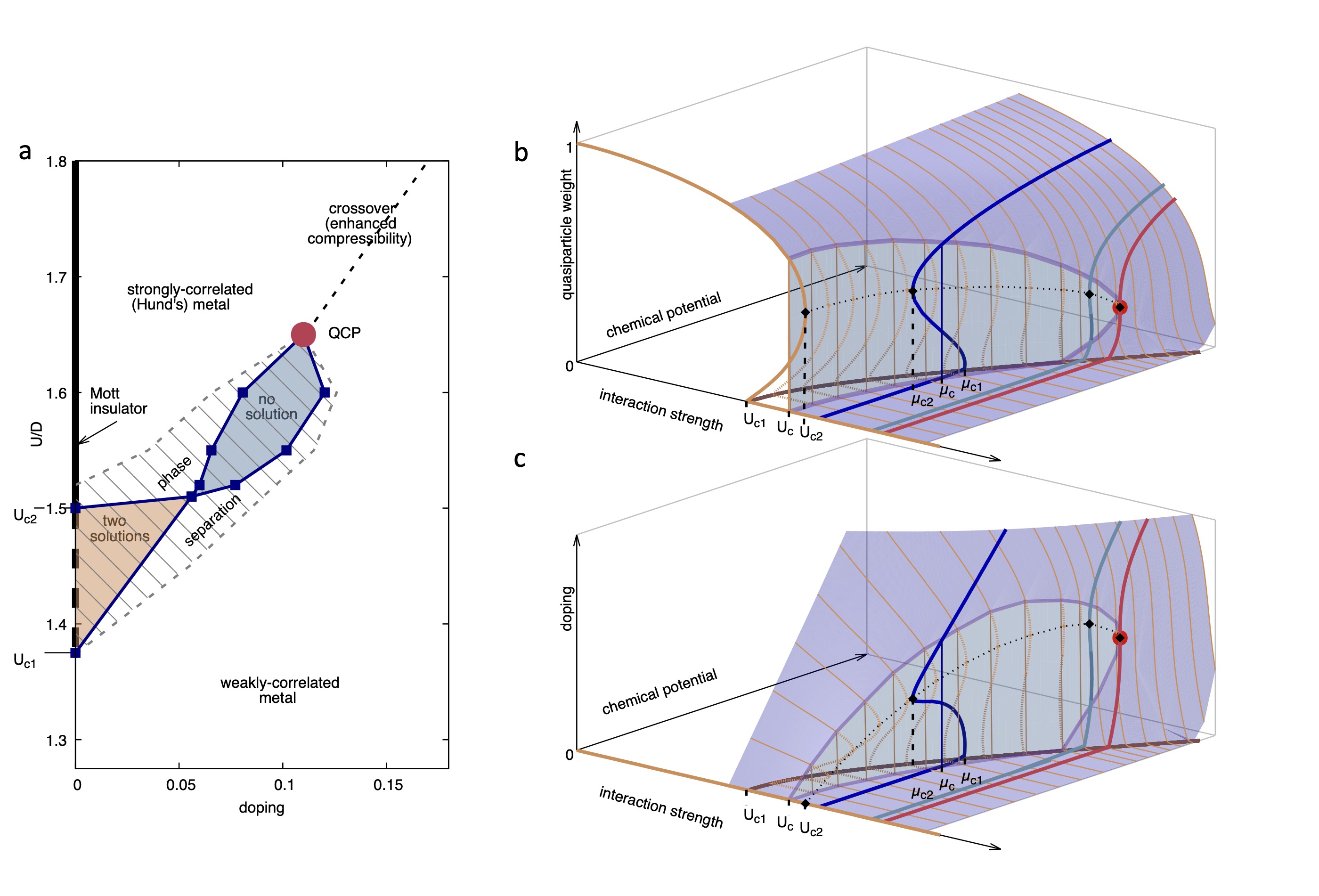}
\caption{{\bf Phase-diagram of the two-orbital Hubbard model and folding of the equation of state. a}, Zero-temperature phase diagram in the interaction strength-doping plane. The thick black line at half-filling is the Mott insulating phase. The dashed black line signals its coexistence with a metal at half-filling. The density-driven Mott transition is first-order for U$_{c1}\!\!\!<$U$\lesssim$U$_{c2}$, due to the first-order nature of the interaction-driven Mott transition, and is accompanied by a zone of phase separation (in this zone, for any given value of U, the stable phase of the system is a mixture of the two homogeneous phases located at the border of the zone for the same U, in proportions needed to obtain the system's doping). For larger interaction strengths the first-order transition is realized between two metals, whereas the density-driven Mott transition becomes second-order. {\bf b}, Illustration of the dependence of the quasiparticle weight Z as a surface plot, as a function of the position in the U-$\mu$ plane. The golden line at $\mu=0$ corresponds to the interaction-driven Mott transition at half-filling (Fig.~\ref{fig:Z_vs_U}a). {\bf c}, Corresponding surface plot of the density (doping from half-filling). Curves are plotted corresponding roughly to those of Fig.~\ref{fig:Z_vs_U}b. The vertical light blue surface corresponds to the Maxwell construction (and physically defines the zone of phase separation) and illustrates the continuity, and common first-order nature, between the interaction-driven Mott transition, the density-driven one in the range near U$_{c1}$ and U$_{c2}$, and the transition between two metals ending in a QCP. This whole scenario is entailed by the folding of the equation of state surface as a multi-valued function on the U-$\mu$ plane (thin golden lines).}\label{fig:PhDiag}
\end{center}
\end{figure*}

These features have substantial consequences for the doped system, which corresponds to a finite chemical potential $\mu$ (our model being half-filled for $\mu=0$). The two solutions evolve differently with $\mu$\cite{Yee_Balents-PhaseSep_Mott}. Indeed as shown in Fig.~\ref{fig:Z_vs_U}b they turn into coexisting stable solutions with different densities $n$ for the same value of $\mu$. Yet they retain their adiabatic connection, giving rise to a sigmoidal shape for the $n(\mu)$ curve, which implies the existence of a zone of phase separation. One can overall visualize (Fig. \ref{fig:PhDiag}b and \ref{fig:PhDiag}c) the equation of state as a surface in three dimensions which is folded in a zone of the U-$\mu$ plane. 

On the other hand, at large doping (large $\mu$) one can expect that all the fingerprints of Mott physics are washed away and a standard metal is recovered. This implies a complete "unfolding" of the equation of state at some $\mu$ after which the system is single valued as a function of thermodynamic parameters (here U and $\mu$). 
Since the paramagnetic Mott insulator can only be realized at integer filling, a possible outcome is that the threshold $\mu$ corresponds to an infinitesimal doping. Our results show instead a different scenario where the two stable solutions survive at finite doping and they all have metallic character. As a consequence, the equation of states unfolds at a finite doping, leading to a finite-doping quantum critical point. 

Calculated $n(\mu)$ curves for several values of U are reported in Fig.~\ref{fig:Z_vs_U}b. For all values U$_{c1}\!\!\!<$U$<$U$_{c2}$ (e.g. for U/D=1.45 - yellow curve in the figure) the two stable solutions existing at half-filling continue at finite $\mu$: the metallic solution is immediately doped while the insulating solution remains pinned at half-filling for a finite range of the chemical potential (corresponding to the gap of the Mott insulator) and eventually becomes doped too. However this doped Mott insulator is a different metal from the continuation of the half-filled metal: both are Fermi liquids (see Supplementary Information) but the former has a much smaller quasiparticle weight, and much lower coherence temperature\cite{Steinbauer_doping-driven_Hund}. 

As we mentioned above, the two stable branches are connected through an unstable solution which implies that the boundaries of the two branches are two spinodal lines where the electronic compressibility  $\kappa=\frac{1}{n^2}\frac{dn}{d\mu}$ diverges.
This also implies the crossing of the free-energies of the two stable solutions at some value of the chemical potential $\mu_c$ in the coexistence range, which then corresponds to a Maxwell construction (see Fig. \ref{fig:PhDiag}b and \ref{fig:PhDiag}c). This determines the first-order nature of the density-driven Mott transition in a range of interactions U$_{c1}\!\!\!<$U$\lesssim$U$_{c2}$, which follows from the first-order nature of the interaction-driven transition and the continuity of the equation of state. The spinodal lines and the approximate phase separation zone are reported in Fig. \ref{fig:PhDiag}a.

For U$>$U$_{c2}$ no metallic solution exists at half-filling and the coexistence zone shifts to larger values of $\mu$, and is seen to shrink (blue curves in Fig.~\ref{fig:Z_vs_U}b). The Maxwell construction jump will then eventually happen at a $\mu_c$ where both branches are metallic.
Therefore the discontinuous Mott transition evolves in a discontinuous transition between two differently doped metals. In contrast, for these (and all larger) values of U, the actual doping-driven Mott metal-insulator transition becomes second order.

Finally, we find that when U grows beyond a critical value U$_{QCP}$ the sigmoid straightens (red curve in Fig.~\ref{fig:Z_vs_U}b), the unstable solution disappears and two stable branches merge into one continuous stable solution. This allows us to establish the existence of a Quantum Critical Point  (QCP) at finite doping, where the two spinodals of the zero-temperature first-order transition merge. There, $\frac{dn}{d\mu}$, thus the electronic compressibility $\kappa$, diverges. 
For U$>$U$_{QCP}$ we are left with a smooth crossover. However the  $\mu$ vs n curve retains an inflection point (black curve in Fig.~\ref{fig:Z_vs_U}b), hence a maximum of the compressibility which culminates in the divergence at the QCP.

This confirms and substantiates the scenario of Ref. \onlinecite{demedici_el_comp,Chatzieleftheriou_RotSym}, where in general a "moustache"-shaped zone of phase separation, delimited by a diverging compressibility and departing from the Mott transition point, crosses the U-doping phase diagram of the Hund's metals.

\begin{figure*}[htb]
\begin{center}
\includegraphics[width=\textwidth]{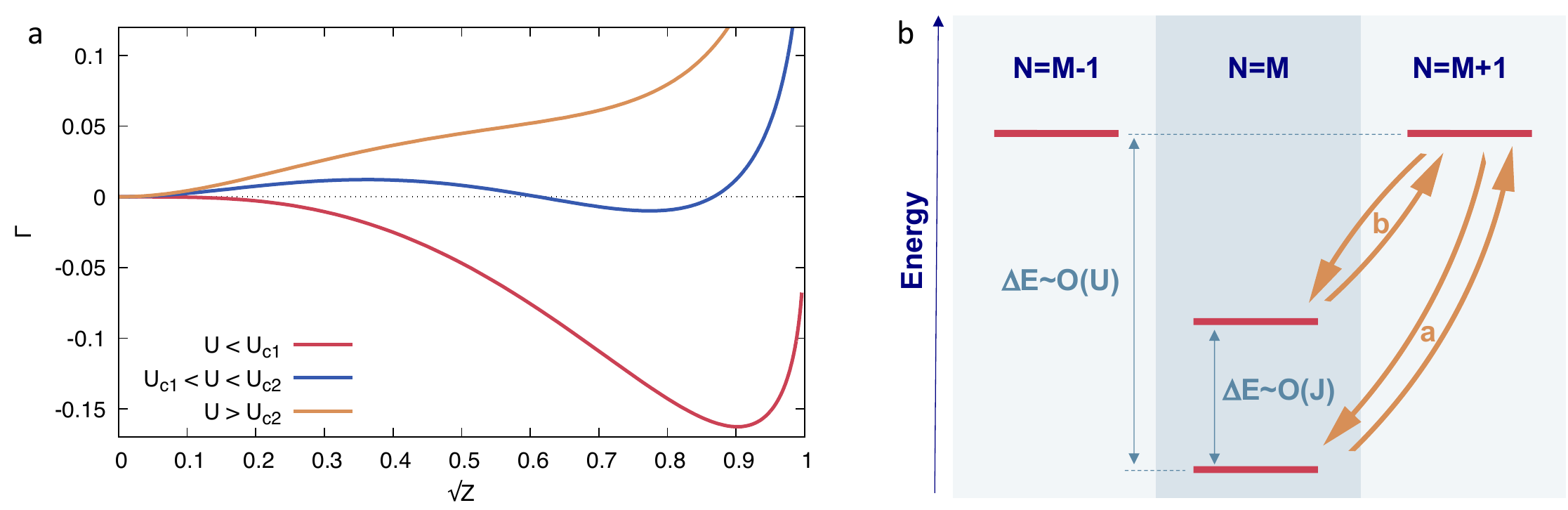}
\caption{{\bf Landay theory of the first-order Mott transition. 
a}, Landau energy function calculated within Slave-Spin mean field (SSMF), where the extrema indicate the stable (minima) and unstable (maxima) equilibrium solutions for the order parameter $m_x/M=\sqrt{Z}$ (where M is the number of orbitals, here M=2, and Z is the quasiparticle weight) for various values of U at half-filling (see Fig. \ref{fig:Z_vs_U}a): below U$_{c1}$ (one minimum at large Z, corresponding to a metal), in the coexistence zone U$_{c1}\!\!\!<$U$<$U$_{c2}$ (two minima - one of which at Z=0, corresponding to the Mott insulator - and one maximum in between), and above $U_{c2}$ (one minimum in Z=0). 
{\bf b}, Sketch of the main features of the atomic spectrum (see Fig.~S7 in the Supplementary Information for the complete spectrum) and hopping processes involved in the strong-coupling perturbation theory giving the fourth-order correction to the ground-state energy $e_4$, which determines the order of the interaction-driven Mott transition. In our model at half-filling the spectrum is symmetric by respect to the half-filled sector (number of electrons N=M). The connected process "a-b-b-a" is negative in sign and its denominator involves a small energy difference $\sim O(J)$ due to the visiting of an excited state in the N=M sector, and thus dominates over the positive lower-order processes (back and forth on the "a" arrows) whose denominators are $\sim O(U)$. This entails a negative $e_4$ and in turn a negative fourth order coefficient $\g_4$ in the Landau energy (see text), thus a first-order Mott transition.}\label{fig:Landau_atomic}
\end{center}
\end{figure*}
It was also shown earlier in Ref.~\cite{Yee_Balents-PhaseSep_Mott} that the incompressible insulating solution emerging from a first-order Mott transition generally crosses its energy again with the coexisting metallic solution at some finite $\mu$,
implying a first-order density-driven Mott transition and phase separation.
We crucially modify and extend this picture here with the continuity between the two solutions, which implies the existence of spinodals limiting the coexistence zone and the phase separation to a small range of U, and with the unfolding of the continuous solution leading to a QCP. 

As a matter of fact, the model we solved realizes a zero-temperature analog of the liquid-gas transition, the role of temperature and pressure being played here by the chemical potential and the interaction strength. 

However this scenario holds only at finite Hund's coupling. At J=0 the transition becomes everywhere second order at zero temperature, and no QCP is realized at finite doping, similarly to the single-band case\cite{Kotliar_SahanaMurthy_compress}.
The key point allowing for the present scenario of phase separation culminating in a QCP is thus the sigmoidal form of the doping-vs-$\mu$ curve at T=0, which is ultimately caused by the first-order character of the interaction-driven Mott transition at half-filling. A natural question thus is: why does the onset of Hund's coupling cause this transition to become first order? How general is the mechanism?


We can get this insight through the analysis of the present model in the Slave-Spin Mean-Field approximation (SSMF - see Methods and  Supplementary Information)\cite{demedici_Vietri}, which is similar to DMFT, but yields simplified yet reliable physics, and here analytically tractable. This method describes the system as a Fermi liquid with quasiparticle weight Z computed from an auxiliary system of quantum spins on a lattice, where it is proportional to the square of the x component of their total on-site magnetization, $m_x$.
The Mott transition maps then onto a ferromagnetic-to-paramagnetic transition of the auxiliary system where $\sqrt{Z}\propto m_x$ plays the role of the order parameter.

The structure of the competing solutions can be analyzed within a Landau theory, by coupling a fictitious external magnetic field $h_{ext}$ conjugated to $m_x$.
The behavior of the numerically calculated Landau energy function (see Supplementary Information) $\Gamma(m_x)$  (Fig.~\ref{fig:Landau_atomic}a) clearly illustrates the first order nature of the transition.

We can calculate $\Gamma(m_x)$ analytically in the vicinity of a Mott insulator, that is around $m_x=0$, where $\Gamma(m_x)= \g_2 m_x^2 + \g_4 m_x^4 + O(m_x^6)$. In order to have the double-minimum structure needed for a first-order transition $\g_4$ has to be negative when $\g_2$ goes from negative to positive for increasing U (which then marks $U_{c1}$ in our case).
We do find that $\g_4 < 0$ for every J$<$U in this model, and we can attach a physical meaning to this result.

Indeed it can be easily shown (see Supplementary Information) that $\gamma_4$ has the same sign of the coefficient $e_4$ in the perturbative expansion of the slave-spin ground state energy $E= e_2 \tilde h^2 + e_4 \tilde h^4 +O (\tilde h^6)$, in terms of the field (self-consistent + external) $\tilde h$ acting on each spin.
In absence of the perturbation ($\tilde h=0$), the slave-spin Hamiltonian reproduces the local atomic spectrum (Fig.~\ref{fig:Landau_atomic}b).
In the half-filled sector (in which the electrons are N=M=2, M being the number of orbitals) the states are split by the Hund's coupling J, while their distance with the sectors with N=3 and N=1 is (U+J)/2. The perturbation changes the occupation, so $e_4$ is due to all the processes involving four hops between neighboring sectors. These processes can be connected (four consecutive jumps starting from and ending in the ground state, but not going through it otherwise - in Fig.\ref{fig:Landau_atomic} these are indicated by the arrows in the order "a-b-b-a"), or disconnected (products of lower-order processes - a roundtrip on the "a" arrows), weighted at the denominator by the energy distance of each intermediate state from the ground state. The connected contributions are always negative while the others are positive at the fourth order.
Now, in the present case only the connected contributions can visit the excited states in the N=2 sector, at distance J in energy from the ground state, thus involving a small denominator. Taking into account all the possible processes they turn out to dominate on the disconnected contributions that involve only the larger energy difference $\sim O(U)$. This causes $e_4$ and thus $\gamma_4$ to be negative.

Hence the ultimate cause of a first-order Mott transition is a splitting of the atomic ground-state multiplet much smaller than the energy cost of charge excitations\cite{FacioCornaglia-Mott_1st_2nd_Order}. 
We can also argue that essentially any term breaking the SU(2M) symmetry leads to the same picture, including, e.g. a Jahn-Teller distortion\cite{Capone_NegativeJ_2band} or a crystal field splitting. 


The scenario linking a first-order Mott transition with phase separation and quantum criticality naturally calls for a connection with models for the cuprates.  In that context, phase separation appears ubiquitously in two-dimensional strongly-correlated models\cite{Emery_Kivelson-PhaseSep_tJ,Grilli_RCDK-PhaseSep_pdmodel,Imada_2D_Cuprate_QMC}. In particular Cluster Dynamical Mean-Field Theory (CDMFT) studies have shown an enhancement of the compressibility at finite temperature culminating with an instability zone which marks the entrance into the pseudogap phase\cite{Sordi_two_metals_CTQMC}. This finite-doping instability causes a first-order transition between two metals across a frontier which can be tracked back to the Mott transition at half-filling, in close analogy with the present analysis. We can thus speculate that this zone ends in a QCP at a critical value of the interaction\cite{Galanakis_Jarrell-QCP_2DHubbard} providing us with a straightforward scenario that connects the indubitable Mott physics with the very existence of a QCP. 

\medskip 

{\noindent \bf  Methods}

{\small
\noindent The model we analyze is the degenerate two-orbital Hubbard model in the paramagnetic phase. The Hamiltonian reads:
\be
\hat H = \sum_{i\neq jm\s} t_{ij} d^\dag_{im\s}d_{jm\s}+\sum_i \hat H^i_{int},
\label{eq:H0}
\ee
where $d^\dag_{im\s}$ creates an electron with spin $\s$ in orbital $m=1,2$ on site $i$ of the lattice, and
\begin{eqnarray}\label{eq:Hint_dens-dens} 
\hat H^i_{int}\,&&=U\sum_{m} \tilde n_{im\up} \tilde n_{im\down}\, +\,U^\prime\!\!\sum_{m\neq m'} \tilde n_{im\up} \tilde n_{im'\down}\,\nonumber \\
&&+(U^\prime-J) \!\!\!\! \sum_{m< m',\sigma} \!\! \tilde n_{im\s} \tilde n_{im'\s} 
\end{eqnarray} 
is the on-site interaction. Here $\tilde n_{im\s}=n_{im\s}-1/2$ is a particle-hole symmetric form of the density operators $n_{im\s}=d^\dag_{im\s}d_{im\s}$, U is the on-site intra-orbital Coulomb repulsion and J the Hund's exchange coupling. We take $J/U=0.25$ and customarily we do not include the off-diagonal terms of the interaction. 
Our analysis depends on the local many-body physics and not on details of the bandstructure obtained by diagonalizing the one-body part of the Hamiltonian $\hat H_0 -\mu \hat N= \sum_{km\s}( \eps_k-\mu) d^\dag_{km\s}d_{km\s}$, where $\mu$ is the chemical potential and $\hat N=\sum_{im\s} n_{im\s}$ is the operator counting the total number of particles in the system. Thus we can choose without loss of generality a featureless bandstructure with semi-circular density of states (DOS) $D(\eps)=2\sqrt{1-(\eps/D)^2}/(\pi D)$ of half-bandwidth D (corresponding e.g. to a Bethe lattice with infinite connectivity). With this choice the model is half-filled (i.e. the electron density $n\equiv\sum_{m\s}\langle n_{im\s}\rangle=2$) for $\mu=0$.

We study this lattice model at T=0 in proximity of half-filling within Dynamical Mean-Field Theory\cite{georges_RMP_dmft}, which yields all the local observables and correlation functions, and in particular the local spectral function $A(\omega)$ and self-energy $\Sigma(\omega)$ of the lattice model. We use several solvers of the DMFT equations: our main results are traced using the Numerical Renormalization Group (NRG). In particular the quasiparticle weight is calculated from the self-energy as $Z=1/(1-\partial{Re\Sigma(0)}/{\partial \omega})$.
Results from other solvers, Exact Diagonalization (ED) and Continuous-Time Quantum MonteCarlo (CTQMC) are discussed in the Supplementary Information and validate mutually with those from NRG.

The Slave-Spin Mean Field approximation\cite{demedici_Vietri} describes the system eqs. (\ref{eq:H0}) and (\ref{eq:Hint_dens-dens}) at half-filling as one of non-interacting fermions with bandwidth renormalized by the quasiparticle weight Z. This factor is computed from an auxiliary system of quantum spins on a lattice (one per orbital $m$ and per spin $\s$ on each site) in the Weiss mean-field approach, and is proportional to the square of the x component of their total on-site magnetization $m_x$:  $\sqrt{Z}\equiv 2\langle S^x_{m\s}\rangle$. The latter is calculated with the slave-spin single-site Hamiltonian $H_s=h^{sc}\sum_{m\s}S^x_{m\s}+H_{int}[S^z]$, where $H_{int}[S^z]$ is the local interaction (\ref{eq:Hint_dens-dens}) with the occupations expressed in the slave-spin space (where a spin "up" corresponds to an occupied fermionic state and "down" to an empty one, so that in the Hamiltonian each $\tilde n_{m\s}$ is replaced with the corresponding $S^z_{m\s}$). 
The self-consistent Weiss field $h^{sc}=8\eps_0\langle S^x_{m\s}\rangle$ flips these occupations embodying the effect of the original hopping term, through $\eps_0=\int_{-\infty}^\mu d\eps D(\eps)\eps<0$ which is the bare kinetic energy of the non-interacting fermions.
}


\acknowledgements The authors are grateful to Rok Zitko for insightful exchanges and to Jakob Steinbauer for sharing some of his data. LdM thanks J. Lorenzana, M. Schir\`o for useful discussions.
MCh, and LdM are supported by the European Commission through the ERC-CoG2016, StrongCoPhy4Energy, GA No724177.
AK and GS were supported by the Deutsche Forschungsgemeinschaft (DFG, German Research Foundation) through Project-ID 258499086 -- SFB 1170 and through the W\"urzburg-Dresden Cluster of Excellence on Complexity and Topology in Quantum Matter ct.qmat (Project-ID 390858490, EXC 2147) and gratefully acknowledge the Gauss Centre for Supercomputing e. V. (www.gauss-centre.eu) for funding this project by providing computing time on the GCS Supercomputer SuperMUC-NG at Leibniz Supercomputing Centre (www.lrz.de).
MB and MC acknowledge support of italian MIUR through PRIN 2015 (Prot. 2015C5SEJJ001) and PRIN 2017 CenTral. 

\bibliographystyle{naturemag}
\bibliography{../Bib/bibldm,../Bib/FeSC,../Bib/hund,../Bib/Janus,../Bib/publdm,../Bib/biblio}

\end{document}


\title{\bf Supplementary Information for: Mott quantum critical points at finite doping}

\author{Maria Chatzieleftheriou}
\affiliation{Laboratoire de Physique et Etude des Mat\'eriaux, UMR8213 CNRS/ESPCI/UPMC, Paris, France}

\author{Alexander Kowalski}
\affiliation{Institut f\"ur Theoretische Physik und Astrophysik and W\"urzburg-Dresden Cluster of Excellence ct.qmat, Universit\"at W\"urzburg, 97074 W\"urzburg, Germany}

\author{Maja Berovi\'{c}}
\affiliation{International School for Advanced Studies (SISSA), Via Bonomea 265, I-34136 Trieste, Italy}

\author{Adriano Amaricci}
\affiliation{CNR-IOM DEMOCRITOS, Istituto Officina dei Materiali,
Consiglio Nazionale delle Ricerche, Via Bonomea 265, I-34136 Trieste, Italy}

\author{Massimo Capone}
\affiliation{International School for Advanced Studies (SISSA), Via Bonomea 265, I-34136 Trieste, Italy}
\affiliation{CNR-IOM DEMOCRITOS, Istituto Officina dei Materiali,
Consiglio Nazionale delle Ricerche, Via Bonomea 265, I-34136 Trieste, Italy}

\author{Lorenzo De~Leo}
\affiliation{Laboratoire de Physique et Etude des Mat\'eriaux, UMR8213 CNRS/ESPCI/UPMC, Paris, France}

\author{Giorgio Sangiovanni}
\affiliation{Institut f\"ur Theoretische Physik und Astrophysik and W\"urzburg-Dresden Cluster of Excellence ct.qmat, Universit\"at W\"urzburg, 97074 W\"urzburg, Germany}

\author{Luca de'~Medici}
\affiliation{Laboratoire de Physique et Etude des Mat\'eriaux, UMR8213 CNRS/ESPCI/UPMC, Paris, France}


\maketitle

\renewcommand{\thepage}{S\arabic{page}}  
\renewcommand{\thesection}{S\arabic{section}}   
\renewcommand{\thetable}{S\arabic{table}}   
\renewcommand{\thefigure}{S\arabic{figure}}
\renewcommand{\theequation}{S\arabic{equation}}


\section{Methods:Dynamical Mean-Field Theory} \label{sec:DMFT}

Dynamical mean-field theory (DMFT)\cite{georges_RMP_dmft} allows to calculate the local correlation functions of the lattice model (eq. (1) of the Methods section) by solving a two-orbital Anderson Impurity Model (AIM) with a suitable bath. Indeed in this AIM, electrons (created by $d_{0m\s}^\dagger$) of a two-orbital impurity having the same local interaction (eq. (2) of the Methods section) as the original lattice model, hop with amplitude $v_l$ from and into a "bath" of non-interacting states of energy $\eps_l$, so that its Hamiltonian reads:
\begin{eqnarray}
\hat H_{AIM}&&=\hat H_{int}[0]-\mu\sum_{m\s}d^\dag_{0m\s}d_{0m\s}\\
&&+\sum_{lm\s} v_l (c_{lm\s}^\dag d_{0m\s})+H.c.)+\sum_{lm\s} \eps_{l} c_{lm\s}^\dag c_{lm\s}.\nonumber 
\end{eqnarray}

The bath is determined by asking that it respects the implicit condition (here in a form specific to the present case of semi-circular DOS) $\Delta(z)\equiv\sum_{l} \frac{{v_l}^2}{z-\eps_l}=\frac{D^2}{4}G_R(z), \forall m,\s$, where $G_R(z)$ is the Fourier transform continued in the plane of complex frequency $z$ of the impurity retarded Green function $G_R(t)=-\theta(t)\langle\{d_{0m\s}(t),d_{0m\s}^\dagger(0)\}\rangle$. This condition is enforced through an iterative self-consistency cycle, and at convergence of this cycle the impurity Green function and self-energy $\Sigma(z)=z+\mu-\sum_{l} \frac{{v_l}^2}{z-\eps_l}-G_R(z)^{-1}$ coincide with the analogous local functions of the lattice model in the DMFT approximation. In particular the local spectral function of the lattice model is given by $A(\omega)=-\frac{1}{\pi}\Im{G_R(\omega+i0^+)}$.

\begin{figure}[t]
\begin{center}
\includegraphics[width=0.5\textwidth]{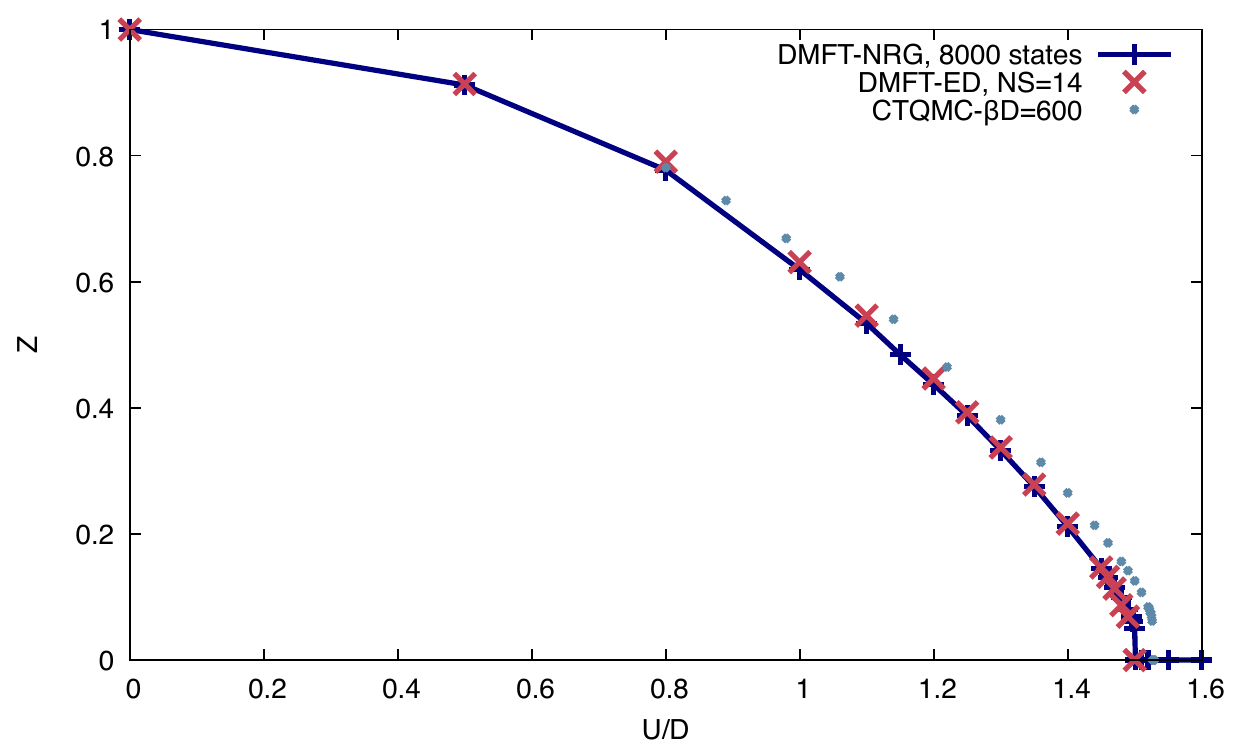}
\caption{Quasiparticle weight Z of the 2-orbital Hubbard model at half-filling, as a function of the interaction strength (density-density interaction and J/U=0.25) calculated within DMFT with the three impurity solvers we use in this article: Numerical Renormalization Group (NRG), Exact Diagonalization (ED) and Continuous-Time Quantum Monte Carlo (CTQMC). The three solvers describe a first-order Mott transition for $U_c \sim1.5D$ (see main text).}\label{fig:DMFT_comparison}
\end{center}
\end{figure}

We solve the AIM with different methods, and in particular we address the zero-temperature properties of the system with the Numerical Renormalization Group (NRG)\cite{nrg} and Exact Diagonalization (ED)\cite{Caffarel_Krauth}, and the finite-temperature ones with Continuous-Time Quantum Monte Carlo (CTQMC)\cite{Gull_Werner-CTQMC_RMP,Wallerberger_Sangiovanni-W2dynamics}, as impurity solvers. 
All these methods have strengths and limitations, but the proximity of their results (Fig. \ref{fig:DMFT_comparison}) validates them mutually. 

Indeed NRG yields the exact low-energy physics in the limit of zero temperature, but describes the spectral features of increasing energy with a decreasing accuracy. Since DMFT connects the different energy scales a quantitative inaccuracy can be present in the final low-energy results which are however qualitatively exact at zero temperature. NRG calculations were performed using a modified version of the NRG Ljubljana code\cite{Zitko_Pruschke_NRG} with quantum numbers corresponding to (Q, S$^z$, T$^z$) which implements a symmetry (U(1)$_{charge}\times$U(1)$_{spin}\times$O(2)$_{orbital}$). The Full Density Matrix algorithm \cite{Weichselbaum_vonDelft_FDM_NRG} was used with $\Lambda$=4.0, 8000 states kept in every NRG iteration, 8 values for the z-interleaving parameter, and a log-normal broadening of the spectral functions with broadening parameter 0.3.
The self-energy was calculated with the so-called Sigma-trick introduced in Ref. \onlinecite{Bulla_Pruschke_NRG}. The quasiparticle weight is evaluated from the real-axis self-energy: $Z=(1-\partial Re\Sigma(0)/\partial \omega)^{-1}$.

ED is the diagonalization of the Hamiltonian of an AIM with a discretized bath (where $l=1,\ldots N_l$), which becomes exact in the limit $N_l\rightarrow\infty$. In practice the method is limited to small $N_l$ by the exponential growth of the Hilbert space to be diagonalized, and this truncation comes with a limit in the spectral resolution and a systematic error. We use this method here to describe the zero-temperature physics in a numerically less expensive framework than NRG. 

At half-filling we can use $N_l=14$ thanks to the Lanczos/Arnoldi algorithm and the new parallel implementation EDIpack\cite{Amaricci_edipack}. The self-consistency equation is enforced on a grid of Matsubara frequencies $\omega_n=\Pi (2n+1)/\beta_{grid}$ with $\beta_{grid}=200/D$, with a weight $1/\omega_n$ in the least-square fit of the discretized variational form of $\D(\omega_n)$ on the numerically obtained $G_R(\omega_n)D^2/4$. The quasiparticle weight is evaluated from the Self-energy on the Matsubara axis: $Z=(1-Im\Sigma(\omega_0)/\omega_0)^{-1}$.
This allows us to cross-validate with NRG the first-order nature of the Mott transition at half-filling and the position of its critical interaction strength. 

Out of half-filling (and in particular in the Hund's metal regime) the Lanczos/Arnoldi algorithm fails (probably due to unresolved degeneracies in the eigenvalues)\cite{Chatzieleftheriou_PhD} and we have to use the full diagonalization of the Hamiltonian matrix, which is limited to much smaller bath sizes. We use then $N_l=6$ and $\beta_{grid}=100/D$.

\begin{figure}[htb]
\begin{center}
\includegraphics[width=0.5\textwidth]{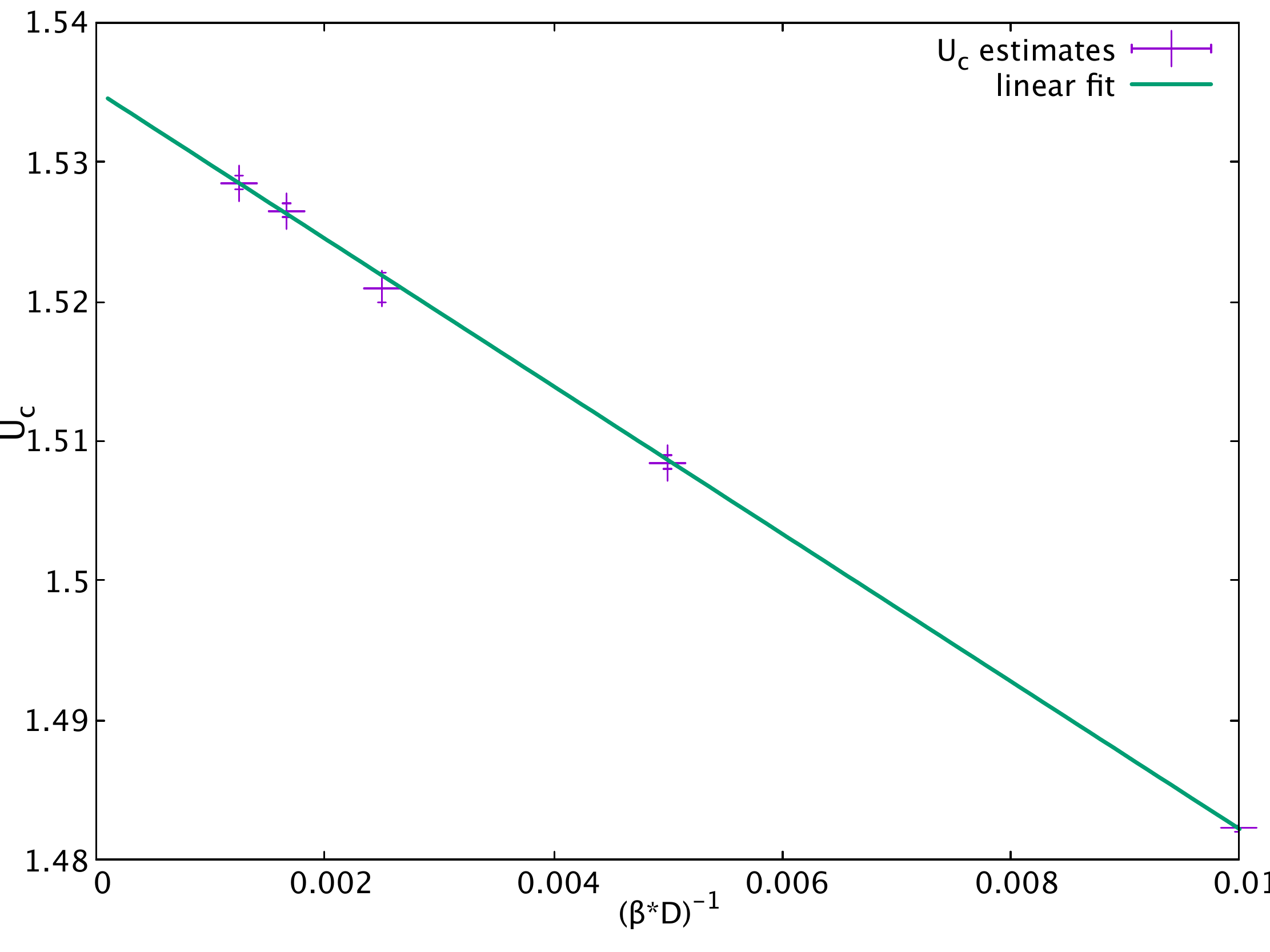}
\caption{Temperature scaling of the critical interaction strength for the Mott transition calculated within DMFT with CTQMC as impurity solver. A linear scaling as a function of the inverse temperature of the values obtained for $\beta D$=100, 200, 400, 600, 800 yields U$_c$/D=1.5351$\pm 0.0004$.}\label{fig:CTQMC_extrap_T0}
\end{center}
\end{figure}
CTQMC\cite{Gull_Werner-CTQMC_RMP} is numerically exact but limited to finite temperature. 
Here we use the CT-HYB solver implemented as part of the w2dynamics package\cite{Wallerberger_Sangiovanni-W2dynamics}.
Around the critical value of the interaction strength, at least 100 DMFT iterations per data point were done regardless of the temperature where in each of the 20 last iterations, a minimum total number of measurement cycles of 300000 was used.
The quasiparticle weight is determined as a $Z=(1-b)^{-1}$ where b is the linear coefficient of a polynomial of degree 2 best-fitted to $Im\Sigma(\omega_n)$ on the lowest 5 Matsubara frequencies.
Here we use CTQMC mainly to obtain low-temperature results that validate the NRG and ED against their possible systematic inaccuracies. 
Indeed in Fig.~\ref{fig:CTQMC_extrap_T0} the critical interaction strength for the Mott transition is estimated at T=0 through temperature scaling using runs at $\beta D$=100, 200, 400, 600, 800.
It is found that the exact value for the half-filled two-orbital model with density-density interaction and J/U=0.25 is U$_c$/D=1.5351$\pm 0.0004$. Thus ED and NRG with the specifications above yield U$_c$/D=1.50, exact within an error of 2$\div$3\%. 
We note good agreement of our QMC results for U$_c$ with those of Steinbauer et al.\cite{Steinbauer_doping-driven_Hund} in cases where data for the same parameters were available for comparison.

\section{NRG numerical evidence of the Fermi-liquid nature of the metallic solutions}

\begin{figure}[t]
\begin{center}
\includegraphics[width=0.5\textwidth]{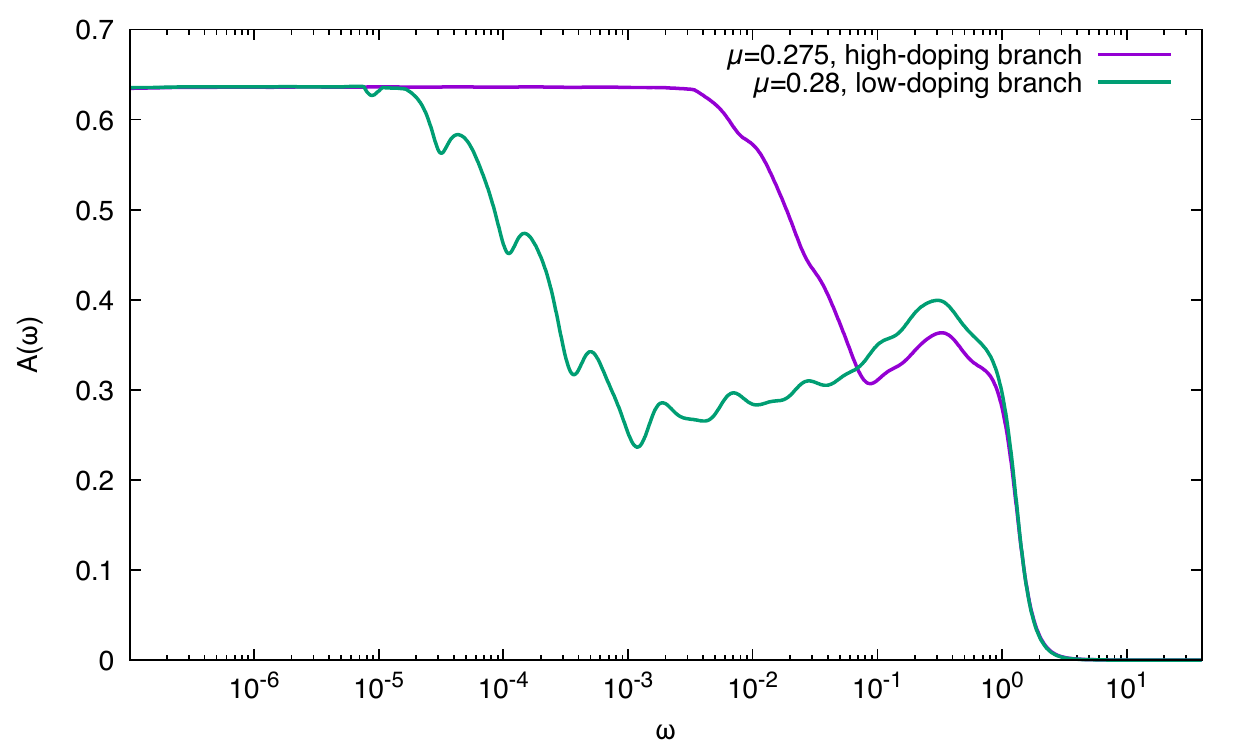}
\caption{Two representative metallic solutions near the phase separation zone, for U=1.6. The Fermi-liquid behaviour of both is signaled by the saturation of the Kondo resonance in the local spectral function (here plotted only at positive frequencies in logarithmic scale) at the pinning value at low frequencies.}\label{fig:NRG_FL}
\end{center}
\end{figure}

At T=0 single-site DMFT is bound to give a Fermi-liquid solution in the single-band Hubbard model\cite{Nozieres_local-FL} and in the multi-band model with positive Hund's coupling\cite{leo04thesis}, if a metal is realized.
It is indeed the case across the whole phase diagram we have described in this work. In the Hund's metal phase however the Fermi-liquid coherence temperature (the Kondo temperature of the AIM solved in DMFT) is found to decrease exponentially fast with decreasing doping\cite{Stadler_Thesis} so that in order to uncover it the resolution of NRG has to be pushed to correspondingly low energies. We illustrate this in Fig. \ref{fig:NRG_FL}, where we show the spectral functions of two stable metallic solutions near the phase separation zone, one on either side of it, for the same interaction U/D=1.6. They have quite different values of doping but approximately the same value of the chemical potential, accordingly indeed to the sigmoidal shape of the n($\mu$) curve (compare to Fig. 1b of the main text). 
The common pinning value reached below some (very different) typical frequency proves the common Fermi-liquid nature (with very different coherence scales however\cite{Steinbauer_doping-driven_Hund}).
This result connects with the proposal of a spin-freezing quantum phase transition in a three-orbital model\cite{Werner_SpinFreezing} associated with a sharp change in the metallic properties across the frontier departing from the half-filling Mott transition. The spin-freezing phenomenon is however limited to finite temperatures, and the spin-frozen phase is replaced by a Fermi liquid at low T\cite{leo04thesis,Georges_annrev}. Our results show that a sharp change of behaviour is already present at zero temperature within the Fermi-liquid phase. This becomes the spin-freezing coherence-incoherence crossover at finite temperatures due to the very different coherence temperature of the two metals. Moreover in this work we have shown that the zero-temperature crossover becomes an actual first-order transition near half-filling.

\section{Compressibility at zero temperature: DMFT-ED}

\begin{figure}[h!]
\begin{center}
\includegraphics[width=0.5\textwidth]{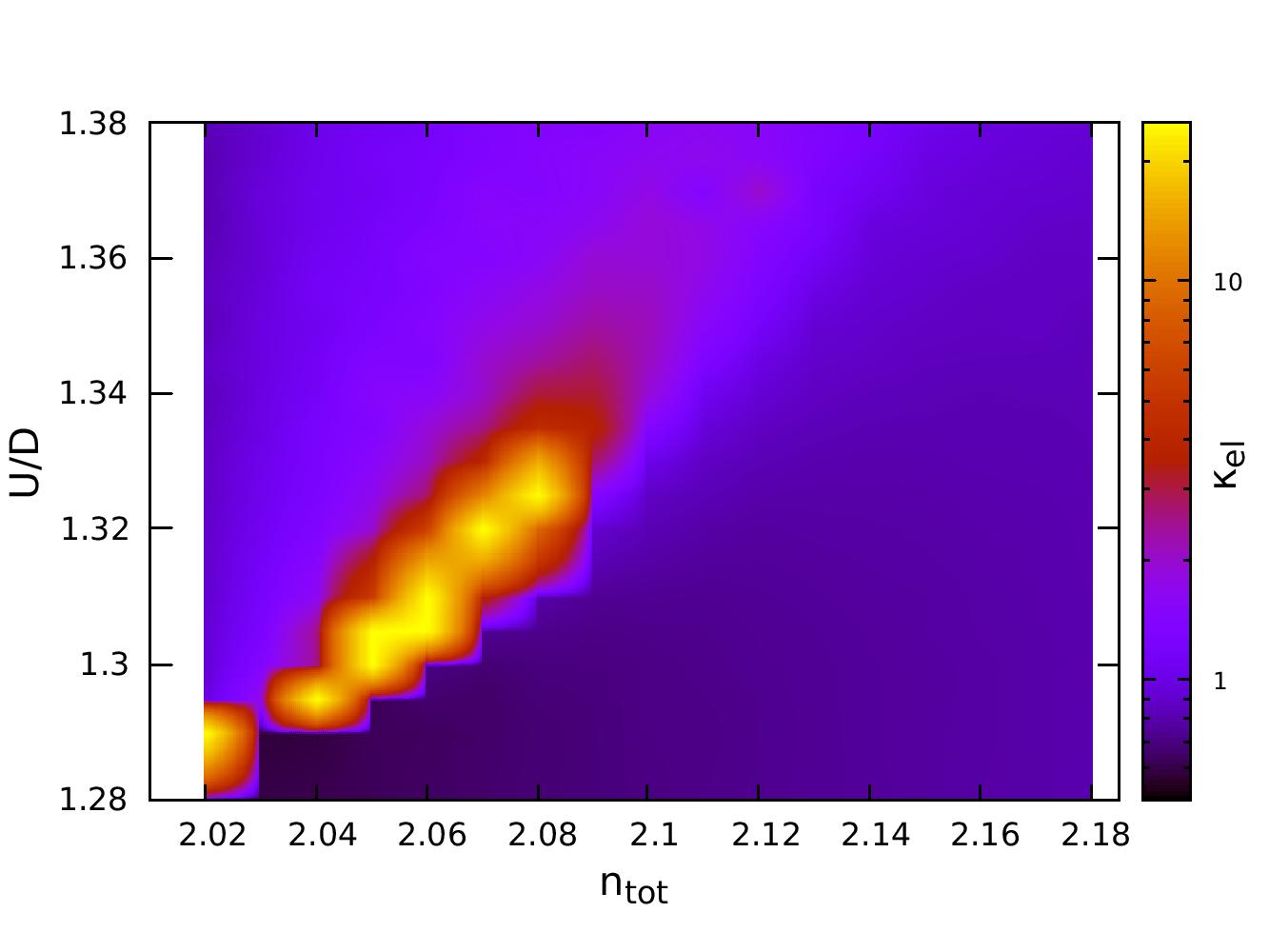}
\caption{Compressibility (color scale) as a function of density and interaction strength calculated by ED ($N_l=6$ and $\beta_{grid}D=100$ - see discussion in \ref{sec:DMFT}). The saturated yellow marks a divergent/negative compressibility (phase separation zone) where the shades of red signal an enhancement tracking the cross-over from the normal (large doping/low U) to the Hund metal.}\label{fig:comp_ED_T0}
\end{center}
\end{figure}

The relative agility of the ED solver for DMFT allows us to trace a complete map of the compressibility enhancement at T=0 in the present model.
This is reported in Fig. \ref{fig:comp_ED_T0}, where the color scale shows how even beyond the quantum critical point (QCP - located here at lower value of doping and U compared to the NRG phase diagram, due to the approximate nature of the ED solver - see the relative information in section \ref{sec:DMFT}) there is a zone (shades of red) where the peak in the compressibility keeps marking the cross over associated to the Hund metal frontier.

\section{Adiabatic continuity of the solutions off half-filling}

In DMFT it is possible to follow the unstable branches which connect the stable solutions coexisting for a given value of U and $\mu$ \cite{Moeller_Dobrosavljevic_unstable_RKKY,Ono_multiorb_linearizedDMFT,Strand_fixpoint_DMFT,Tong_Hubbard_sigmoid_1storder}. At half-filling we have calculated, with NRG as an impurity solver, the corresponding reentrant behaviour of the quasiparticle weight Z as a function of the interaction strength U/D at T=0, as displayed in Fig.~1a of the main text. 
\begin{figure}[h!]
\begin{center}
\includegraphics[width=0.5\textwidth]{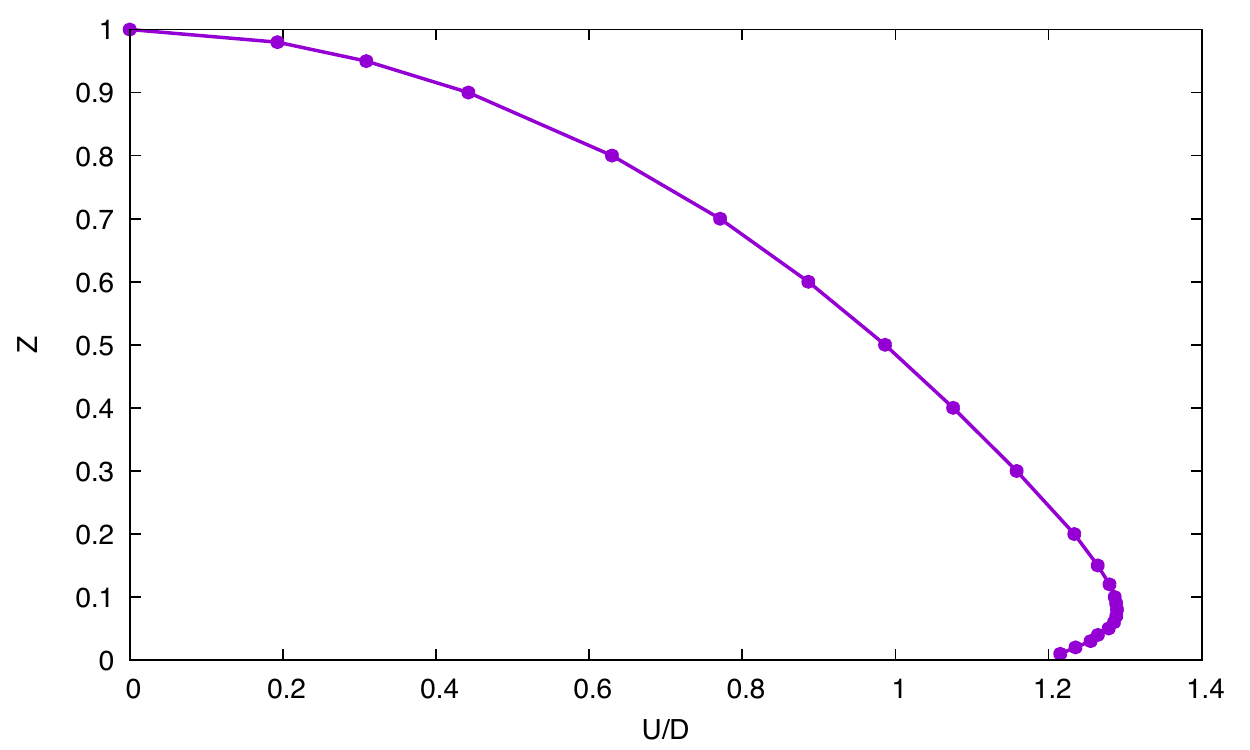}
\includegraphics[width=0.5\textwidth]{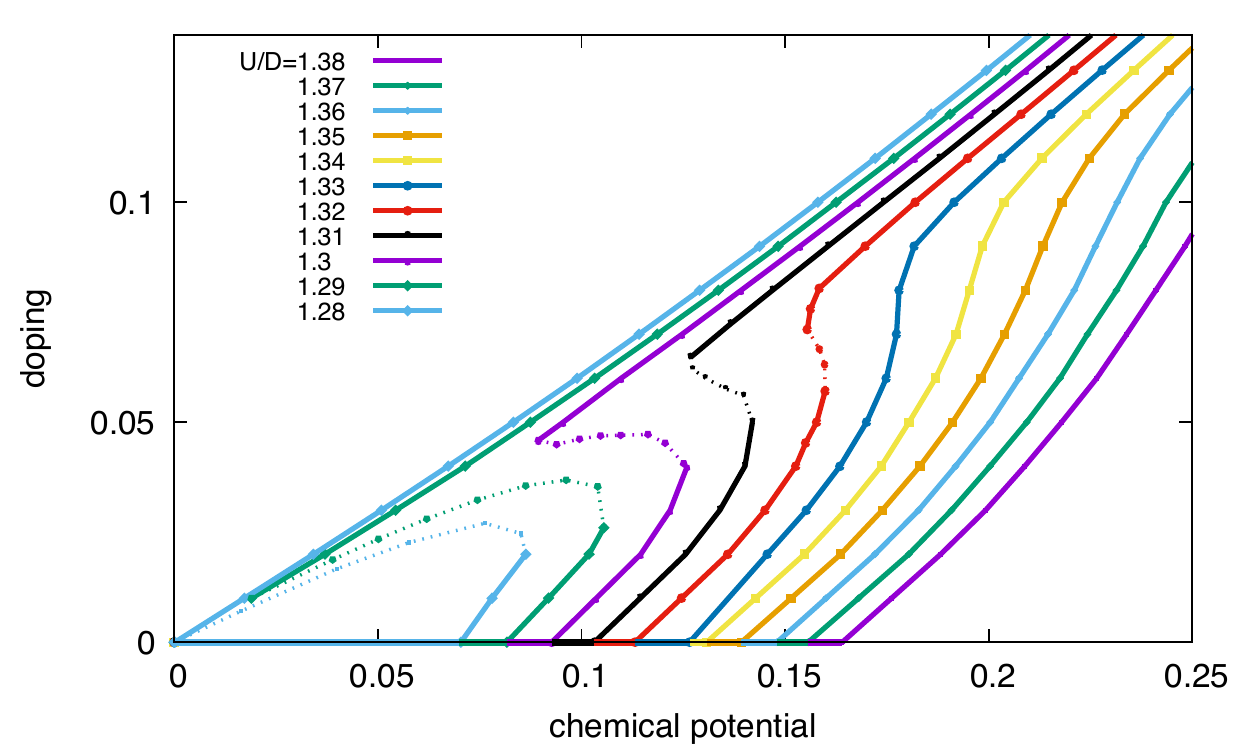}
\caption{DMFT results with Exact Diagonalization as an impurity solver with $N_l$=6 at T=0, showing the sigmoidal behaviour of the quasiparticle weight Z as a function of the interaction strenght U/D at half-filling (upper panel, see Fig.~1a of the main text) and of the density as a function of the chemical potential off half-filling (lower panel, see Fig.~1b of the main text). These data illustrate the adiabatic connection of the stable branches of coexisting solutions through unstable ones, which holds everywhere in the phase diagram of the present model solved with DMFT, enabling us to discuss it in terms of the "folding" of the equation of state surface. The results in the lower panel are used to sketch the unstable branches in Fig.~1b of the main text, where only the stable ones are data calculated with NRG.}\label{fig:adiab_ED}
\end{center}
\end{figure}

In this section we report the analogous results from ED at T=0. We use here $N_l$=6, which ensures a semi-quantitative agreement with the more exact NRG and QMC calculations, but can be used both at and off half-filling (see the corresponding discussion in section \ref{sec:DMFT}). 
In the upper panel of Fig. \ref{fig:adiab_ED} the Z(U) plot shows indeed a behaviour analogous to that of Fig.~1a of the main text, only slightly shifted to lower values of U/D owing to the aforementioned approximation. In the lower panel we report the calculated density vs $\mu$ curves, showing indeed the sigmoidal behaviour corresponding to Fig.~1b of the main text. 

In all these calculation a parameter $\alpha$ ($\alpha=$U at half-filling, $\alpha=\mu$ off half-filling) is adjusted at every DMFT iteration in order to search for a wanted value of $Z=Z_{target}$. The latter is indeed found to monotonically decrease in all these curves, thus it labels uniquely a point along the curves, unlike the physical parameters as a function of which these curves are indeed multi-valued functions. We use, at iteration i+1, $\alpha(i+1)=\alpha(i)+(Z(i)-Z_{target})\chi(i)$ and the step is $\chi(i)=(\alpha(i)-\alpha(i-1))/(Z(i)-Z(i-1))$.
While the stable solutions can be found also in the standard way by fixing the physical parameters and using a suitable guess (typically the previously converged solution on the same branch) to start the iterative DMFT loop, a converged solution in an unstable branch is reached only using this protocol. We consider converged one such solution not only when the calculated and the target Z differ by less than a given low threshold, but also when the changes in the variable $\a$ at each iteration $|\a(i)-\a(i-1)|$ satisfy an analogous criterion. Obviously also the usual DMFT criterion of convergence is enforced, which requests that the local Green functions (or equivalently the baths $\D$ or the self-energies) calculated at iterations $i$ and $i-1$ (summed in absolute value over the frequency grid) differ by less then a given threshold.   
This requires typically a considerably larger number of iterations compared to the stable branches. This is also why we have studied in detail the unstable branches off half-filling only with ED, and then sketched the corresponding result on Fig.~1b of the main text, where only the points on the stable branches are actually calculated with NRG.

\section{Methods:Slave-spin Mean-Field and its perturbative expansion: coefficients in the Landau theory of the Mott transition}

In the Slave-Spin Mean-Field (SSMF) method\cite{demedici_Slave-spins,demedici_Vietri} the lattice system to be studied (eq. (1) of the main text) is replaced by an analog in an enlarged Hilbert space, where for each fermionic degree of freedom of the original model there is a fermion with the same indices (here called $f_{im\s}$) and an accompanying "slave" quantum spin-1/2 of components $S^{x,y,z}_{im\s}$.

An occupied local fermionic state of the original model is represented by the state $|1\rangle_f|1\rangle_S$, and the corresponding unoccupied state by $|0\rangle_f|0\rangle_S$, where $|1\rangle_S$ and $|0\rangle_S$ represent the states "up" and "down" of the z-axis projection of the slave spin $S^z$, respectively. These "physical" states are such that $n^f_{im\s}= S^z_{im\s} + \frac{1}{2}$, and this condition distinguishes them from the "unphysical" states $|1\rangle_f|0\rangle_S$ and $|0\rangle_f|1\rangle_S$, i.e. which do not represent any state of the original system, and should be excluded from the averages yielding the physical quantities.
This cannot be done rigorously and in practice averages are performed on the whole enlarged Hilbert space and this condition is enforced through Lagrange multipliers guaranteeing only that $\langle n^f_{im\s}\rangle=\langle S^z_{im\s}\rangle + \frac{1}{2}$.

The Hamiltonian operator has then to be expressed in the new Hilbert space. The hopping terms thus, besides moving the fermions from one site to another also flip the corresponding spins (through the $S^\pm_{m\s}$ operators).  The density-density interaction instead can be expressed only in terms of the slave-spins' $S^z_{im\s}$  z-components only. 
Then the slave-spins and the fermions are mean-field decoupled and the lattice slave-spin Hamiltonian is further treated in a Weiss single-site mean-field. In the end the problem is approximated by $\hat{H}-\mu N=\hat{H}_{f}+\hat{H}_{s}$, a fermionic Hamiltonian of independent quasiparticles $\hat{H}_{f}$ self-consistenly coupled to a collection of single-site Hamiltonians with 2M (where M is the number of orbitals, here M=2) slave-spins interacting on-site $\hat H_s=\sum_{i} \hat H^i_s$.
These equations are solved numerically in an iterative fashion. Further details on the method can be found in Ref. \onlinecite{demedici_Vietri} and in the appendix A of Ref. \onlinecite{Chatzieleftheriou_RotSym}.

At half-filling for the present degenerate, particle-hole symmetric system the SSMF equations take the simple form (the Lagrange multipliers vanish by symmetry, and $\mu$=0):

\begin{equation}
\hat{H}_{f}=\sum_{i\neq j,m\sigma}Zt_{ij}f_{im\sigma}^{\dagger}f_{jm\sigma},
\end{equation}
where the renormalization factor 
\begin{equation}\label{eq:Z}
Z= 4 \langle S^x_{im\s}\rangle^2,
\end{equation}
is calculated through the single-site slave-spin Hamiltonian (dropping the site index i) and:
\begin{equation}\label{eq:SS}
\hat{H}_{s}=h^{sc}\sum_{m\sigma}S^x_{m\sigma}+H_{int}
\end{equation}
\begin{align}  \label{eq:Hint}
H_{int}=&U\sum_{m}S_{m\uparrow}^{z}S_{m\downarrow}^{z}+U'\sum_{m\neq m'}S_{m\uparrow}^{z}S_{m'\downarrow}^{z} \nonumber \\  
+&(U'-J)\sum_{m<m'\sigma}S_{m\sigma}^{z}S_{m'\sigma}^{z}
\end{align}

The self-consistent Weiss field reads:
\begin{equation}
h^{sc}= 8\langle S^x_{m\s}\rangle \eps_0,\\
\end{equation}
where 
\begin{equation}
\eps_0\equiv \sum_{j\neq i} t_{ij}\langle f_{im\s}^\+ f_{jm\s}\rangle=\int_{-\infty}^\mu d\eps D(\eps)\eps
\end{equation}
is the kinetic energy of the non-interacting system. At particle-hole symmetry ($\mu=0$, independently of U, implying half-filling) this is a constant, and the self-consistency has to be enforced only within the spin Hamiltonian.

\begin{figure}[t]
\begin{center}
\includegraphics[width=0.5\textwidth]{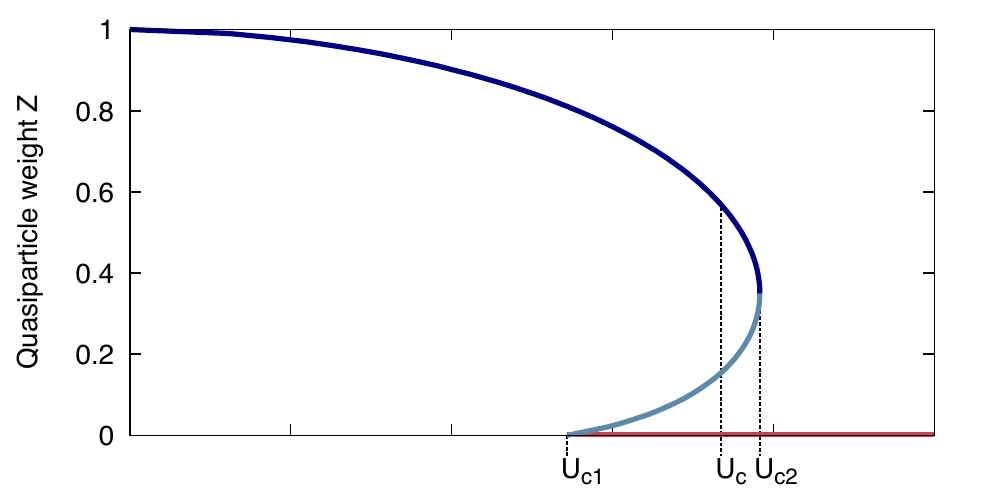}
\includegraphics[width=0.5\textwidth]{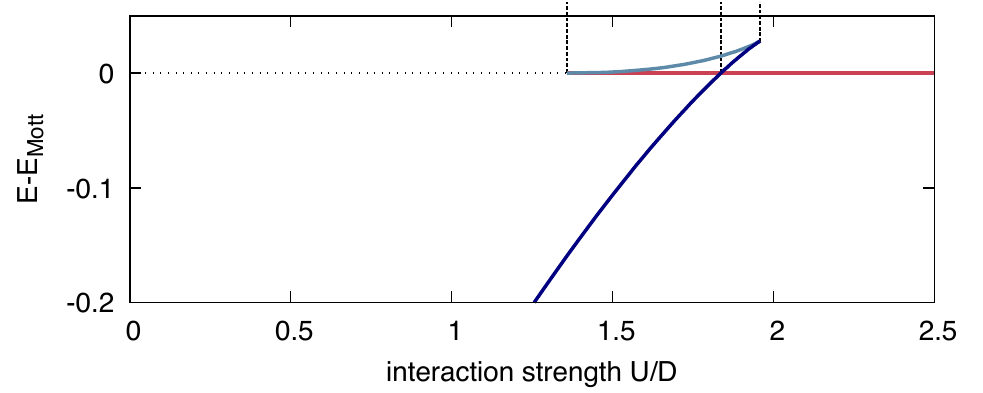}
\caption{Upper panel: quasiparticle weight for the half-filled 2-orbital Hubbard model with density-density interaction and J/U=0.25 calculated within SSMF. Three solutions (one stable metallic, one stable insulating, connected by an unstable metallic branch) indeed coexist in the range U$_{c1}\leq$U$\leq$U$_{c2}$, analogously to the DMFT solution reported in Fig. 1a of the main text. The interaction-driven Mott transition is thus first-order, as implied by the crossing of the energy of the two stable solutions at U$_c$, reported in the lower panel.}\label{fig:Z_vs_U_SSMF_2orb}
\end{center}
\end{figure}

A vanishing Z indeed signals the Mott transition. The SSMF result for the 2-band Hubbard model with J/U=0.25 is shown in Fig. \ref{fig:Z_vs_U_SSMF_2orb}. The sigmoidal shape of the Z vs U curve clearly signals the first-order nature of the transition.

The expression for Z eq. (\ref{eq:Z}) shows that Mott transition in SSMF is described as the vanishing of the x-component of the slave-spin on-site magnetization $m_x\equiv2M\langle S^x_{im\s}\rangle$. It is thus mapped onto a ferromagnetic-paramagnetic transition of the auxiliary spin system, and $\sqrt{Z}\sim m_x$ plays the role of an order parameter.
 
\emph{Landau theory for the Mott transition.} 
The stability and competition between the solutions around this transition can be studied in the framework of a Landau theory.
 A fictitious external magnetic field $h_{ext}$ conjugated to $m_x$ can be introduced through a linear term in the Hamiltonian proportional to $\sum_{m\s} S^x_{m\s}$, turning eq. (\ref{eq:SS}) into $\hat{H}_{s}=\tilde h\sum_{m\sigma}S^x_{m\sigma}+H_{int}$, where $\tilde h\equiv h^{sc}+h_{ext}$.

The corresponding Gibbs free energy\footnote{All energies are intended per site.} is $\Gamma(m_x)=E(h_{ext}(m_x))-h_{ext}(m_x)m_x$, where $E(h_{ext})=\langle H_s+h_{ext}\sum_{m\s}S^x_{m\s} \rangle$ is the energy of the ground state obtained in presence of the field $h_{ext}$. 
In this mean-field approximation $\Gamma(m_x)$ can then be viewed as a Landau function\cite{negele}: $\frac{\partial \Gamma}{\partial m_x}=-h_{ext}$ so that the extrema of $\Gamma(m_x)$ indicate the equilibrium solutions in absence of driving field.
In Fig. 3a of the main text it is illustrated how the minima in $\Gamma(m_x)$ (calculated numerically for a range of finite $h_{ext}$) describe the Mott transition and the characteristic double-minimum structure of its first order nature in the present system with nonzero J. Indeed in the proximity of the Mott state ($m_x=0$) the Landau function can be expanded: 
\be
\Gamma(m_x)= \g_2 m_x^2 + \g_4 m_x^4 + O(m_x^6).
\ee 
In order to have the double-minimum structure needed for a first-order transition, $\g_4$ has to be negative when $\g_2$ changes sign from negative to positive for increasing U (which then marks $U_{c1}$ in our case).

We can calculate these coefficients analytically using standard perturbation theory.
The implied relationship between field and magnetization is 
\be\label{hext_vs_m_1}
h_{ext}=-\frac{\partial \Gamma}{\partial m_x}=-2\g_2 m_x - 4\g_4 m_x^3 + O(m_x^5),
\ee
the coefficients of which can be estimated by explicitly calculating $m_x(h_{ext})$ in our model and inverting it.

Indeed we can compute the groundstate energy of  $H_s$ up to the fourth order in the total field $\tilde h=h_{ext}+h^{sc}$ 
(the external field plus the self-consistent one $h^{sc}=4\eps_0m_x/M$) 
felt by each slave-spin, through standard perturbation theory: 
\be
E= e_2 \tilde h^2 + e_4 \tilde h^4 +O (\tilde h^6).
\ee 
This implies $m_x=\frac{\partial E}{\partial \tilde h}=2e_2 \tilde h + 4e_4 \tilde h^3+O(\tilde h^5)$, which can be inverted in $\tilde h=1/(2e_2)m_x-4e_4/(2e_2)^4 m_x^3$. 
Subtracting the self-consistent field one finally gets:
\be\label{hext_vs_m_2} 
h_{ext}=(1/(2e_2)-4\eps_0/M)m_x-4e_4/(2e_2)^4 m_x^3.
\ee
Comparing eq. (\ref{hext_vs_m_2}) with eq. (\ref{hext_vs_m_1}) shows that $\gamma_4$ has the same sign of $e_4$, which we can calculate.

\begin{figure}[t]
\begin{center}
\includegraphics[width=0.5\textwidth]{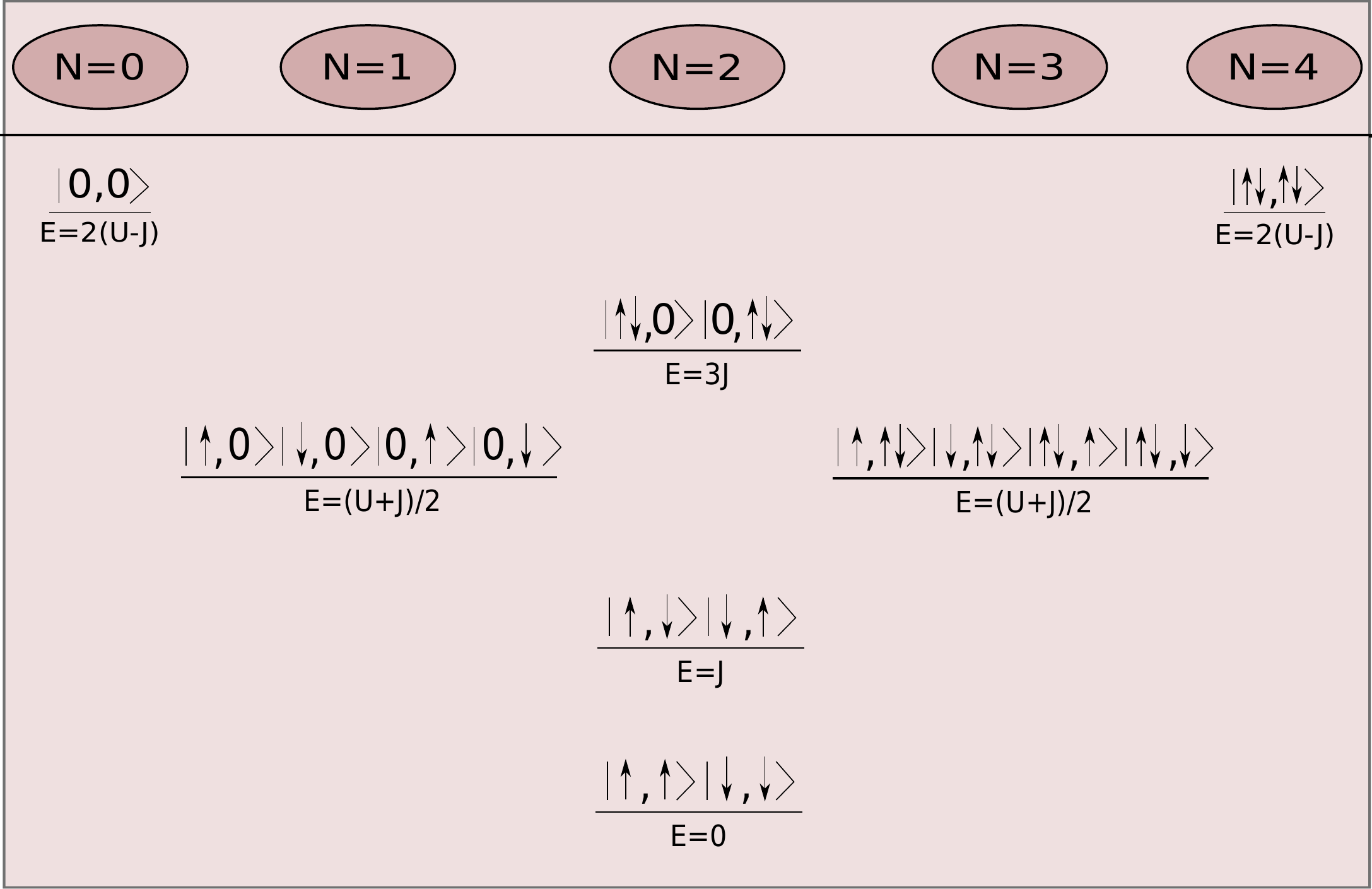}
\caption{Spectrum of $H_{int}$ eq. (\ref{fig:spectrum}), used in our perturbative analysis. For compactness of notation, the slave-spin states are here indicated with the ket of the physical state they represent. }\label{fig:spectrum}
\end{center}
\end{figure}

\emph{Perturbative expansion of the Slave-Spin ground-state energy.}
Let's then solve the slave-spin equations in proximity of the Mott insulator $Z=4\langle S^x_{m\s}\rangle=0$, which implies $h^{sc}=0$.
In absence of field, for $\tilde h=0$, the spectrum is that of eq (\ref{eq:Hint}), illustrated in Fig. \ref{fig:spectrum} (we recall that we take $U'=U-2J$).

For small $\tilde h$ we can apply perturbation theory, the operator of the perturbation being $V\equiv \tilde h \sum_{m\sigma}S_{m\sigma}^{x}$, and the nonvanishing terms of the second order and fourth order corrections to the energy of the ground state $|\phi\rangle$ read:
\begin{align}\label{eq:E_pert}
&E_{\phi}^{(2)}=\frac{|V_{\phi\kappa_{2}}|^{2}}{E_{\phi\kappa_{2}}}, \\
&E_{\phi}^{(4)}=\frac{V_{\phi\kappa_{4}}V_{\kappa_{4}\kappa_{3}}V_{\kappa_{3}\kappa_{2}}V_{\kappa_{2}\phi}}{E_{\phi\kappa_{2}}E_{\phi\kappa_{3}}E_{\phi\kappa_4}}-E_{\phi}^{(2)}\frac{|V_{\phi\kappa_{4}}|^{2}}{E_{\phi\kappa_{4}}^{2}}\label{eq:fourth}
\end{align}
where we have defined $V_{\mu\kappa}=\langle\mu^{0}|V|\kappa^{0}\rangle$ and $E_{\mu\kappa}=E_{\mu}^{0}-E_{\kappa}^{0}$ (the superscript 0 indicates unperturbed states and energies), and all the $\kappa_i$ are meant summed on all the states such that none of the $E_{\phi\kappa_i}$ vanishes. 

Note that $V_{\phi\phi}=0$ in this case because the eigenstates of the unperturbed Hamiltonian $H_{int}$ are eigenstates of $S_{m\sigma}^{z}$, while the perturbation flips the spins (indeed $S_{m\sigma}^{x}=(S_{m\sigma}^{+}+S_{m\sigma}^{-})/2$). Only neighbouring sectors in Fig. \ref{fig:spectrum} (i.e. sectors with only one particle more or less) are connected through $V$, so the terms including $V_{\phi\kappa_{3}}V_{\kappa_{3}\kappa_{2}}V_{\kappa_{2}\phi}$ (an odd number of "hops" having to start and end in the ground state) vanish as well. The odd-order contributions are made of these terms and thus vanish altogether, and hence $E(\tilde{h})=e_{2}\tilde{h}^{2}+e_{4}\tilde{h}^{4}+O(\tilde{h^{6}})$.

As motivated in the main article we are interested in the sign of the fourth-order contribution of the ground state energy, which determines the order of the Mott transition.
The key observation here is that in the case of the ground state all the $E_{\phi\kappa}<0$, and all numerators are positive. Thus the first term in eq. (\ref{eq:fourth}) (which we call "connected" since it involves a chain of "hops" starting and ending in the ground state, without going through it - or its degenerate manifold - otherwise) is negative, while the second (a product of lower-order, "disconnected", terms) is positive.
Moreover only the connected term can visit the lowest excited states in the same charge sector (N=2) of the ground state (that is, spin excitations here), which lie at energy distance J from it. The lowest order terms (and thus the disconnected contributions involving them) only visit the states at distance $\frac{U+J}{2}\sim O(U)$. For small enough J then, the first term in eq. (\ref{eq:fourth}) will dominate over the second, since its terms have a smaller denominator, and thus the sign of the whole fourth-order contribution will be negative.

 Performing all the summations we obtain: 
\begin{equation}\label{eq:Egs_h}
E(\tilde{h})=-\frac{2}{U(1+j)}\tilde{h}^2-\frac{2(7j^2-9j+8)}{U^{3}(1+j)^{3}(1-j)3j}\tilde{h}^4,
\end{equation}
where $j=J/U$. The numerator of the fourth order coefficient is definite positive so that the whole coefficient $e_4$ is negative as long as of J$<$U.
This implies that $\gamma_4<0$ and the Mott transition is first-order in this model as long as of J$<$U (that is, for all realistic situation in a material).
This analysis substantiates and rationalizes the empirical rule of thumb proposed in Ref. \onlinecite{FacioCornaglia-Mott_1st_2nd_Order}, that "the transition tends to be first order if the lowest-lying excitations are in the same charge sector as the atomic ground state".

This is further confirmed in the analogous model with Kanamori interaction\cite{Georges_annrev} where the splitting of the ground state multiplet is 2J rather than J. There the connected terms will decrease more quickly for increasing J, compared to the present density-density case. Indeed numerical results for the Kanamori case show that the first-order jump at half-filling first increases for increasing J, and then after a maximum it decreases until vanishing, and the transition becomes second-order again\cite{Klejnberg_Spalek_Hund,Lechermann_RISB,demedici_Vietri}.

\section{5-orbital model and relevance for the Iron-based superconductors}

\begin{figure}[t]
\begin{center}
\includegraphics[width=0.5\textwidth]{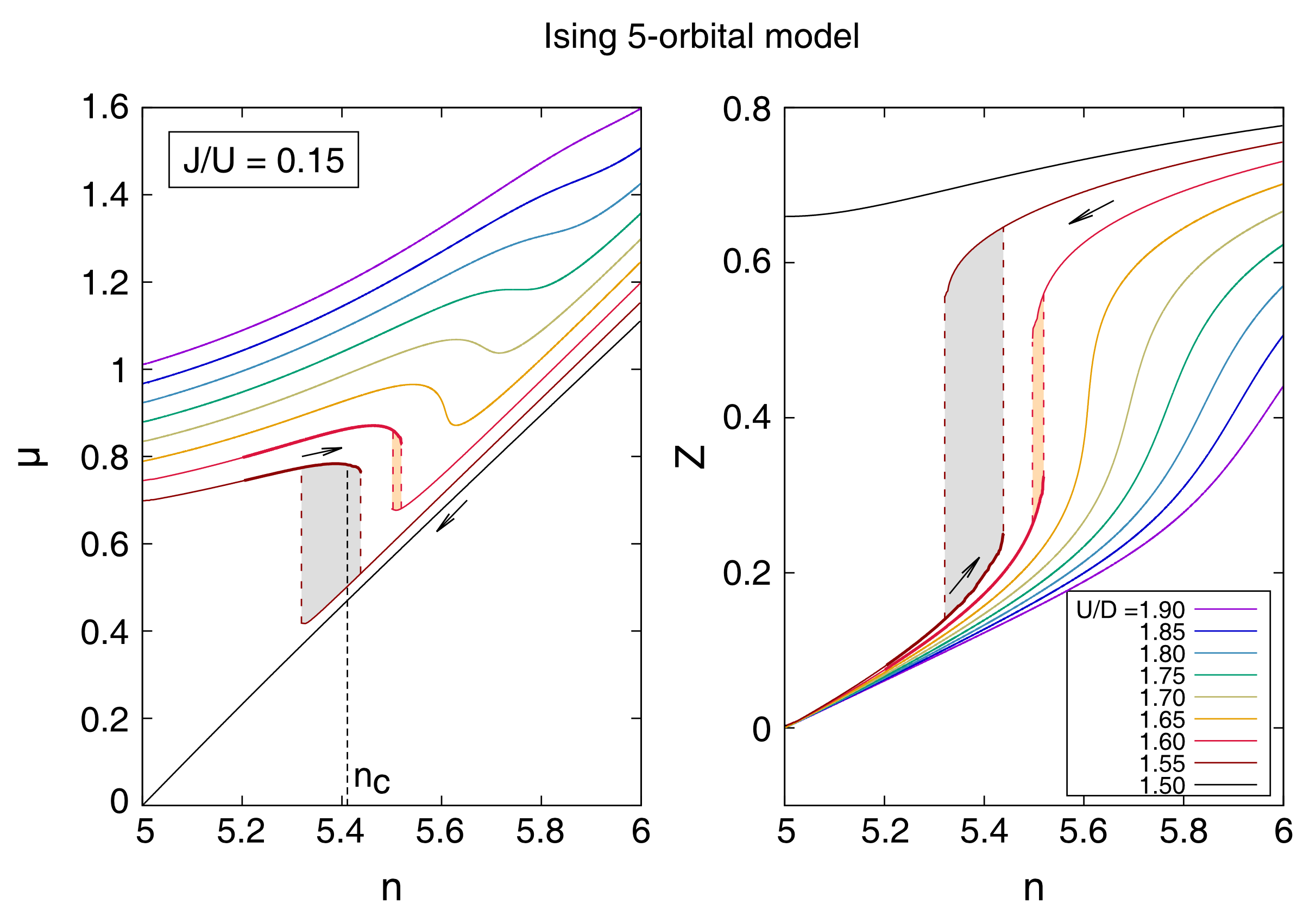}
\caption{Analogous results for the 5-orbital model, calculated with SBMF. Again both the bi-stability and the phase separation are realized, ending in a QCP. The rightmost panel illustrates the strong difference in the degree of metallicity between the stable (Hund) metal at low doping and the conventional one realized at large doping.}\label{fig:5-orbital}
\end{center}
\end{figure}

An analogous phenomenology is displayed by Hubbard models with more than 2 orbitals, as partly explored in Ref. \onlinecite{Chatzieleftheriou_RotSym}. Here we highlight the presence of the QCP ending the phase separation zone. Indeed the $\mu$ vs density curves calculated within Kotliar-Ruckenstein Slave-Boson Mean-Field  (SBMF, which yield results similar, and in certain cases identical to SSMF) for a 5-orbital model are shown in Fig. \ref{fig:5-orbital}.  The phase separation zone, like in the 2-orbital case is characterized by the sigmoidal behaviour, with several coexisting solutions for a given chemical potential in a range of values of U. We also highlight how - like in the 2-orbital case - upon reducing U towards U$_{c1}$ the two stable branches can also overlap in doping yielding a "bi-stability" zone (indicated as "two solutions" in Fig. 2a of the main text).  The further inflection of the curve due to the multiple solutions at a given doping evolves in the coexistence of metallic and insulating solutions at half-filling. This confirms in the generic M-orbital case the result of the main article, that the adiabatic connection of the solutions connects the first-order nature of the Mott transition at half-filling with the phase separation zone off-half-filling, ending in a QCP at finite doping.

\bibliography{../Bib/bibldm,../Bib/FeSC,../Bib/hund,../Bib/Janus,../Bib/publdm,../Bib/biblio}